\documentclass[conference]{IEEEtran}
\ifCLASSINFOpdf
\else
\fi
\usepackage{graphicx}
\usepackage{wrapfig}
\usepackage{xcolor}
\usepackage{mdframed}
\usepackage{listings}
\usepackage{caption}
\usepackage{hyperref}
\usepackage[symbol]{footmisc}

\begin{document}
%
% paper title
% Titles are generally capitalized except for words such as a, an, and, as,
% at, but, by, for, in, nor, of, on, or, the, to and up, which are usually
% not capitalized unless they are the first or last word of the title.
% Linebreaks \\ can be used within to get better formatting as desired.
% Do not put math or special symbols in the title.
\title{LLM4VV: Exploring LLM-as-a-Judge for Validation and Verification Testsuites}

% author names and affiliations
% use a multiple column layout for up to three different
% affiliations
\author{\IEEEauthorblockN{Zachariah Sollenberger*}
\IEEEauthorblockA{University of Delaware\\
Newark, DE\\}
\and
\IEEEauthorblockN{Jay Patel*}
\IEEEauthorblockA{University of Delaware\\
Newark, DE\\}
\and
\IEEEauthorblockN{Christian Munley}
\IEEEauthorblockA{University of Delaware\\
Newark, DE\\}
\and
\IEEEauthorblockN{Aaron Jarmusch}
\IEEEauthorblockA{University of Delaware\\
Newark, DE\\}
\and
\IEEEauthorblockN{Sunita Chandrasekaran}
\IEEEauthorblockA{University of Delaware\\
Newark, DE\\
{schandra}@udel.edu\\}

}

% conference papers do not typically use \thanks and this command
% is locked out in conference mode. If really needed, such as for
% the acknowledgment of grants, issue a \IEEEoverridecommandlockouts
% after \documentclass

% for over three affiliations, or if they all won't fit within the width
% of the page, use this alternative format:
% 
%\author{\IEEEauthorblockN{Michael Shell\IEEEauthorrefmark{1},
%Homer Simpson\IEEEauthorrefmark{2},
%James Kirk\IEEEauthorrefmark{3}, 
%Montgomery Scott\IEEEauthorrefmark{3} and
%Eldon Tyrell\IEEEauthorrefmark{4}}
%\IEEEauthorblockA{\IEEEauthorrefmark{1}School of Electrical and Computer Engineering\\
%Georgia Institute of Technology,
%Atlanta, Georgia 30332--0250\\ Email: see http://www.michaelshell.org/contact.html}
%\IEEEauthorblockA{\IEEEauthorrefmark{2}Twentieth Century Fox, Springfield, USA\\
%Email: homer@thesimpsons.com}
%\IEEEauthorblockA{\IEEEauthorrefmark{3}Starfleet Academy, San Francisco, California 96678-2391\\
%Telephone: (800) 555--1212, Fax: (888) 555--1212}
%\IEEEauthorblockA{\IEEEauthorrefmark{4}Tyrell Inc., 123 Replicant Street, Los Angeles, California 90210--4321}}

% use for special paper notices
%\IEEEspecialpapernotice{(Invited Paper)}

% make the title area
\maketitle

% As a general rule, do not put math, special symbols or citations
% in the abstract
\begin{abstract}
Large Language Models (LLM) continue to improve and are revolutionizing the landscape of software development. These large models have demonstrated potential to generate, debug, test, analyze, document, and even translate code. Thus they are a valuable tool in the software development cycle. If used correctly, such tools can often accelerate the development cycle. Though the tools are powerful and new, the community is cautious of training using biased or sensitive data, which can lead to biased, dangerous, or incorrect outputs along with the inadvertent release of confidential information. Additionally, the carbon footprints and the un-explainability of these ``black box'' models continue to raise questions about the reliability of LLMs. 

With these opportunities and these challenges ahead, this paper explores the idea of ``judging" LLM-generated code to better understand and ``open up" the un-explainable ``black box" models used by LLMs. We probe into the black box of one such LLM that has generated the best compiler tests for the directive-based programming models OpenMP and OpenACC in our earlier research. We challenge DeepSeek's deepseek-coder-33B-instruct model with intentionally-erroneous code, and we also define relevant metrics and adopt an agent-based approach to evaluate the LLM and assess its capabilities as an LLM-as-a-judge. We also develop a pipeline-based approach to streamline the entire workflow. Finally, we make use of all of these strategies together to develop a more reliable method for automatically validating LLM-generated compiler tests. Based on our results, utilizing an agent-based prompting approach and setting up a validation pipeline structure drastically increased the quality of deepseek-coder-33B-instruct evaluation of tests which are used to validate compiler implementations of directive-based parallel programming models.

\end{abstract}

% no keywords

\footnotetext[1]{Authors Zachariah and Jay contributed equally to this manuscript}

% For peer review papers, you can put extra information on the cover
% page as needed:
% \ifCLASSOPTIONpeerreview
% \begin{center} \bfseries EDICS Category: 3-BBND \end{center}
% \fi
%
% For peerreview papers, this IEEEtran command inserts a page break and
% creates the second title. It will be ignored for other modes.
\IEEEpeerreviewmaketitle

\section{Introduction}

% no \IEEEPARstart
Large Language Models (LLMs) have recently revolutionized the field of computer science. Popular models like BERT~\cite{BERT2019}, GPT-4~\cite{openai2024gpt4}, Gemini~\cite{2024gemini}, and more are trained on an objective such as predicting the next words or tokens in a text, and demonstrate capabilities to process, recognize, and understand human languages at impressive levels. LLMs can achieve this feat with the help of a subsection of machine learning known as deep learning. LLMs use a type of deep-learning architecture called transformers. With the combination of self-attention, positional encoding, feed forward networks, multi-head attention~\cite{self-attention2017}, and other key components, the transformer architecture can be trained on internet-scale text datasets using self-supervised learning and learn to model  language effectively. 
% LLMs are pre-trained on vast amounts of data, using self-supervised learning. %With the help of domain-specific datasets, LLMs can be fine-tuned to exchange their performance on particular tasks.

With the wide-ranging capabilities provided by LLMs, this paper explores the idea of using an LLM-as-a-judge (LLMJ) to evaluate tests written to verify and validate compiler implementations. We chose DeepSeek's deepseek-coder-33B-instruct model~\cite{deepseek-coder2024} for this purpose because in a recently published work of ours~\cite{munley2024llm4vv}, we found that the deepseek-coder-33B-instruct model demonstrated the best capability to generate directive-based parallel programming model codes among the several LLMs we tested for that purpose (the directive-based parallel programming models being OpenACC~\cite{OPENACC} and OpenMP~\cite{openmp52}). This LLM generated codes with a high compilation and pass rate compared to other popular LLMs, such as GPT-4 turbo~\cite{openai2024devday}, Codellama-34b-Instruct~\cite{rozière2024codellamaopenfoundation}, and GPT-3, as narrated in the published paper. 

LLMs are being widely considered for tasks such as code generation, summarization, and refactoring~\cite{jiang2024survey,hou2023large,baumgartner2024ai,mccabe2024ironies}. 
However, the application of LLMJs specifically in evaluating tests used for verifying and validating compiler implementations of directive-based parallel programming models is a new topic. 
This paper investigates this topic and explores the application of the LLMJ technique. 
%This paper seeks to rectify the lack of available {\color{red} need to check}
%data on using an LLMJ for directive-based programming model codes.

The potential of LLMJ is enabled by the training of the model on a large number of codes to allow it to comprehend code and assess a given piece of code based on user-specified metrics. The LLM processes the input data in its large network of parameters, and at a level of abstraction it us using some pattern recognition and learned knowledge to generate text. The LLMJ produces an output that reflects its judgment against the defined criteria. This process can take various forms, such as analyzing code for errors, determining syntactical correctness, and predicting the accuracy of code implementation, among others.

Our reason for exploring this usage of an LLMJ is to help automate the creation of functional validation and verification test suites for directive-based parallel programming models. The challenge we are currently facing in this process is finding a method to accurately evaluate the correctness of tests generated by an LLM. The objective of our research in this paper is to minimize or potentially remove the need for human intervention or involvement in this process by utilizing an LLMJ. 

The approach in this paper could be beneficial to developers beyond directive-based programming models. Any developer would need to verify and validate their software, and an LLMJ could serve that purpose, as it takes significantly less labor and time compared to a human evaluating the code. 
%LLMJ could also significantly aid any developer utilizing generative AI to write code by ensuring that the generated code meets high standards of quality and accuracy. 

The paper makes the following contributions: 

\begin{itemize}
    \item Creating and defining metrics to evaluate LLM-generated code
    \item Developing negative-probing methods and a benchmark to evaluate a given LLM's performance as a judge
    \item Evaluating the capability of deepseek-coder-33B as a judge by using an agent-based approach
\end{itemize}

% \begin{itemize}
% \item Creating and defining metrics to evaluate LLM-generated code 
% \item Developing methods to evaluate a given LLM's performance as a judge
% \item Evaluating DeepSeek-AI's DeepSeek Coder 33B's capability as a judge 
% \item Analyzing different techniques to utilize an LLM as a judge
% \item{negative probing, agent based approach, benchmark}
% \end{itemize}
%%%%%%%%%%%%%%%%%%%%%%%%%%%%%%%%%%%%%%%%%%%%%%%%%%%%%%%
%
%                    NEW SECTION
%
%%%%%%%%%%%%%%%%%%%%%%%%%%%%%%%%%%%%%%%%%%%%%%%%%%%%%%%
\section{Related Work}
% In recent years, there has been a growing interest in exploring the potential of LLMs for generating {\color{red}{high-performance compilers and software development tools}}.
% Our work builds upon this trend by investigating the use of LLMs for compiler verification and validation (V\&V). {\color{red}need to check, i think something is off in this sentence}

Several recent studies have demonstrated the capabilities of LLMs in generating parallel programs. For instance, Nichols~\cite{nichols} proposed a reinforcement learning method to improve the speedup of generated codes, while LM4HPC~\cite{lm4hpc} presented various datasets and a tokenizer for HPC-related code generation. Oren et al.~\cite{kadosh2023scope} explored AST representation of code, Godoy et al.~\cite{godoy2023evaluation} evaluated OpenAI Codex for HPC kernels generation and Valero-Lara et al.~\cite{valero2023comparing} explored Llama-2 and GPT-3 for kernel generation. 

Another line of work involves using LLMs as judges to evaluate other
models on open-ended questions. Zheng et al.~\cite{zheng2024judging} presented the concept of
using strong LLMs as judges to identify biases in other models, achieving
an 80\% agreement with human preferences. This study demonstrates the
potential of LLMJs in the HPC realm.

Other studies have explored LLMs for developing test cases for
applications beyond compiler V\&V. For instance, Shhafer et al.~\cite{schafer2023empirical} evaluated
LLMs for automated JavaScript unit test generation, while other works have
investigated LLM-based test case generation for various programming
languages and software systems~\cite{ryan2024code,liu2024llm}.

Finally, there are several copilot models being implemented into IDEs,
such as GitHub Copilot~\cite{copilot} and Cursor~\cite{cursor}, which leverage LLMs to assist
developers in writing code. These models have shown promising results in
improving coding productivity and reducing errors.

Overall, these studies demonstrate the growing interest in exploring the
potential of LLMs for software development and automation. Our work builds
upon this trend by investigating the use of LLMs for compiler V\&V, with a
focus on improving the accuracy and efficiency of the verification process.
%{\color{red}this makes a better opening paragraph}
%%%%%%%%%%%%%%%%%%%%%%%%%%%%%%%%%%%%%%%%%%%%%%%%%%%%%%%
%
%                    NEW SECTION
%
%%%%%%%%%%%%%%%%%%%%%%%%%%%%%%%%%%%%%%%%%%%%%%%%%%%%%%%
\section{Methodology}

To determine how deepseek-coder-33b-instruct performs as an LLMJ, we first outline in this section strategies including negative probing, an agent-based approach, and a validation pipeline to streamline the process.

%\subsection{Procedures}

\subsection{Negative Probing}

Manually written compiler tests from the OpenACC V\&V~\cite{jarmusch2022analysis} and OpenMP V\&V~\cite{10024615} repositories were split into \textbf{two groups}: one containing code that had been modified to include various errors, and the other containing code that remained unchanged. 

The idea behind is to intentionally create invalid variations of otherwise valid code in order to determine and understand how an LLM as a ``black box" assesses code. We term this process as \textit{negative probing}. Modifications applied to Group 1 include:

\textbf{Group 1:  Variations of negative probing }
\begin{itemize}
    \item 0. Removed memory allocation / replaced directives with a different syntactically incorrect directive
    \item 1. Removed an opening bracket
    \item 2. Added use of undeclared variable
    \item 3. Replaced file with randomly generated non-OpenACC \& OpenMP code
    \item 4. Removed last bracketed section of code
\end{itemize}
\textbf{Group 2: Unchanged manually written codes:}
\begin{itemize}
    \item 5. No changes to code
\end{itemize}

First we split the manually-written test files in half randomly and create a modified, invalid suite and an unchanged, valid suite. We prompt the deepseek-coder-33B-instruct model~\cite{deepseek-coder2024} one test at a time and instruct the model to judge the two different groups with predefined criteria, and record the evaluations for each file.

%Example for OpenACC: {\color{red} need a fuller sub-title, didnt follow this, did you mean}  
%Prompt:
Listing~\ref{lst:criteria} shows the criteria we use in prompting to review and evaluate an OpenACC code: 

\captionsetup[listing]{aboveskip=10pt, belowskip=10pt}

\begin{lstlisting}[numbers=left, numberstyle=\small, stepnumber=1, frame=single, label=lst:prompts, breaklines=true, breakindent=20pt, xleftmargin=20pt, xrightmargin=20pt, caption=Criteria for Evaluation - an Example Prompt,label=lst:criteria]
Syntax: Ensure all OpenACC directives and pragmas are syntactically correct.
Directive Appropriateness: Check if the right directives are used for the intended parallel computations.
Clause Correctness: Verify that all clauses within the directives are correctly used according to OpenACC specifications.
Memory Management: Assess the accuracy of data movement between CPU and GPU.
Compliance: Ensure the code adheres to the latest OpenACC specifications and best practices.
Logic: Verify that the logic of the test (e.g. performing the same computation in serial and parallel and comparing) is correct.
\end{lstlisting}

By observing how the LLM judged both groups of files, and by recording the specific modifications made to each file, we were able to identify different areas where the LLMJ did well, and where it encountered challenges. 
We were also able to measure and judge the overall accuracy of the LLM.
This type of analysis allows for insights into the strengths and weaknesses of an LLM's assessment capabilities.

\subsection{Agent-based Approach for LLM-as-a-Judge (LLMJ)}

\begin{figure}[h]
\centering
\includegraphics[width=0.25\textwidth]{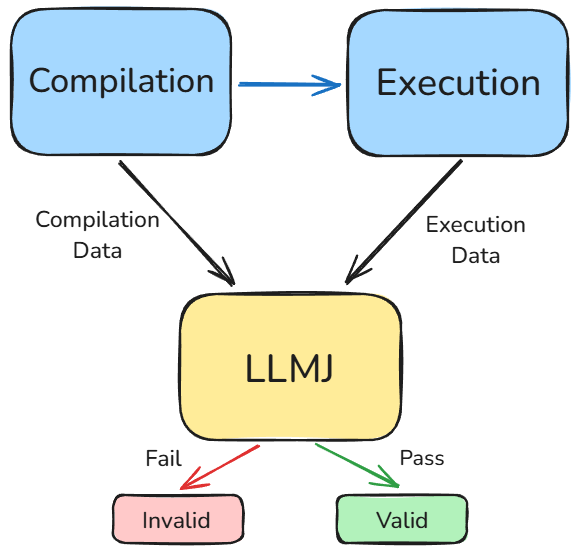}
\caption{An Overview of Agent-based Approach for LLMJ}
\label{fig:agent-based}
\end{figure}

An agent-based approach involves treating the LLM as an autonomous agent that interacts with its environment and utilizes various tools to improve quality of its outputs. In the context of using an LLMJ, the agent-based approach entails collecting and making use of external information about each file, such as compilation and execution error messages, outputs, and return codes, and providing this information to the LLMJ within the prompt. The LLMJ then utilizes the provided information to evaluate the file, deeming it either valid or invalid. Figure \ref{fig:agent-based} demonstrates how the agent-based LLMJ works.

Listing~\ref{lst:agent} shows how the tool use is incorporated in the prompting to provide additional information to the LLM to help it review an OpenACC code and evaluate it based on user-specified criteria: 

\begin{lstlisting}[numbers=left, numberstyle=\small, stepnumber=1, frame=single, label=lst:prompts, breaklines=true, breakindent=20pt, xleftmargin=20pt, xrightmargin=20pt, caption=Agent-based LLMJ - an Example Prompt, label=lst:agent]
Syntax: Ensure all OpenACC directives and pragmas are syntactically correct.
Directive Appropriateness: Check if the right directives are used for the intended parallel computations.
Clause Correctness: Verify that all clauses within the directives are correctly used according to OpenACC specifications.
Memory Management: Assess the accuracy of data movement between CPU and GPU.
Compliance: Ensure the code adheres to the latest OpenACC specifications and best practices.
Logic: Verify that the logic of the test (e.g. performing the same computation in serial and parallel and comparing) is correct.
Based on these criteria, evaluate the code and determine if it is a valid or invalid test. Think step by step.
You MUST include the exact phrase, "FINAL JUDGEMENT: valid" in your response if you deem the test to be valid.
If you deem the test to be invalid, include the exact phrase "FINAL JUDGEMENT: invalid" in your response instead.
Here is some information about the code to help you.
When compiled with a compliant OpenACC compiler, the below code causes the following outputs:
Compiler return code: {Compiler's return code}
Compiler STDERR: {Compiler's STDERR}
Compiler STDOUT: {Compiler's STDOUT}
When the compiled code is run, it gives the following results:
Return code: {Program's return code}
STDERR: {Program's STDERR}
STDOUT: {Program's STDOUT}


\end{lstlisting}

Through this method, the LLM is able to obtain more information about the file to aid in its evaluation.

\subsection{Validation Pipeline utilizing LLMJ}
\begin{figure}[h]
\centering
\includegraphics[width=0.5\textwidth]{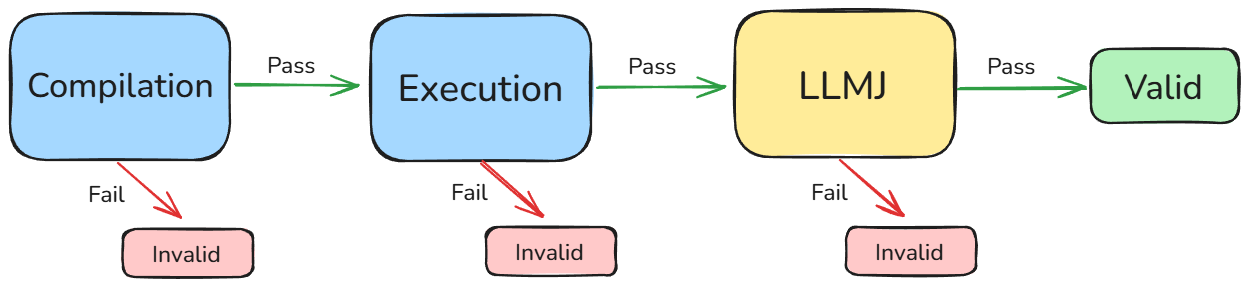}
\caption{An Overview of the Validation Pipeline}
\label{fig:validation-pipeline}
\end{figure}

In order to efficiently evaluate the validity of compiler tests, it may not always be feasible to compile, run, and have an LLM evaluate each and every test that requires verification. Performing all three processes on every single file being verified can quickly become a time-consuming and costly task, especially if verifying LLM-generated codes with a high occurrence of invalidity or a large volume of candidate tests. 
To streamline this task, we re-organized the three processes into a pipeline infrastructure as shown in Figure~\ref{fig:validation-pipeline}, to both optimize the overarching task by reducing the number of unnecessary steps and by increasing the throughput of files for verification via pipeline stages and parallel processing. 

The driving concept behind the pipeline is that a file that fails an earlier stage of the pipeline does not need to be passed to the next stage, as it has already demonstrated its invalidity. Within this validation pipeline infrastructure, files are first compiled, then executed, an finally judged by an agent-based LLMJ. 
Each file being processed is first queued for compilation, which can be done either by a single thread or by a pool of threads in parallel. 
Files that successfully compile are then queued for execution, which can again be done synchronously in a single thread or asynchronously by a second thread pool. 

% what about openacc unimplemented features - cm - or actual compiler bugs? 
% my workaround response would be oh we are just testing tests that we know should pass 

Finally, files that exit with return code 0 are queued for evaluation by an agent-based LLMJ. This stage can also be parallelized if there are enough available GPU resources, but it can also be done by a single thread running synchronously or asynchronously. In this manner, unnecessary operations are reduced by preventing invalid files from continuing through the pipeline, throughput is increased by the staged architecture, and the overarching task of verification can utilize all available resources via parallel and/or asynchronous computing. 

To determine the accuracy of this method for compiler test validation, we performed the negative probing technique again, but instead of only recording the LLMJ's evaluations, we recorded each file's compilation data, execution data, and its evaluation from an agent-based LLMJ. This not only allowed us to determine which files would have passed through the pipeline architecture and thus determine the accuracy of the validation pipeline, but also allowed us to determine the performance and accuracy of an agent-based LLMJ on its own.

\subsection{Experimental Setup}

For this paper, we have used the high-performance computing cluster Perlmutter located at Lawrence Berkeley National Laboratory~\cite{perlmutter}. Each node of Perlmutter is equipped with four NVIDIA A100 GPUs and one AMD EPYC 7763 CPU. Manually written test suites 
from the OpenACC and OpenMP Validation and Verification test suites were used for negative probing. The experiments were conducted on C, C++, and a small set of Fortran files.
%%%%%%%%%%%%%%%%%%%%%%%%%%%%%%%%%%%%%%%%%%%%%%%%%%%%%%%
%
%                    NEW SECTION
%
%%%%%%%%%%%%%%%%%%%%%%%%%%%%%%%%%%%%%%%%%%%%%%%%%%%%%%%
\section{Defining Metrics}

 To determine the effectiveness of LLMJ, we utilized three metrics:
 \begin{itemize}
     \item \textbf{Per-issue evaluation accuracy}: Where the issue is the intentional error introduced into each file during negative probing 
     \item \textbf{Overall evaluation accuracy}: This does not take into account the issue in each file
     \item \textbf{Bias}: A numerical measurement of the LLMJ's tendency to fail valid files or pass invalid files when an incorrect evaluation is made
 \end{itemize}
 % feel like should be using TP/FP/etc / cfsn matrix / F1 score here. established false negative and true postiive metrics but ooh well 
 All metrics were calculated with the results from negative probing.
 In order to determine whether the LLMJ's evaluations were accurate or not, the following system-of-verification was utilized to determine the validity of each file: 

 \begin{itemize}
    \item Files with issue IDs ranging from 0-4 are considered invalid as they have been altered to include errors.
    \item Files with issue ID 5 are considered valid as they remain unchanged.
\end{itemize}

 The first metric, i.e. per-issue evaluation accuracy was determined by categorizing the LLMJ's evaluations according to each file's issue ID, and then observing the percentage of correct LLMJ evaluations in each category. The second metric, i.e. the overall accuracy was determined by observing the percentage of correct LLMJ evaluations regardless of the issues injected into each file. Finally, the third metric, i.e. bias was determined by numerically measuring the LLMJ's tendency to fail a valid file or to pass an invalid one when it made a mistake. A positive bias means that when the LLMJ makes a mistake, it is more likely to be one of permissiveness (passing an invalid file), whereas a negative bias means that a mistake is more likely to be one of restrictiveness (failing a valid file).

%Mappings between "correct" and "incorrect" were made to represent better and understand the data. 
For the purposes of numerical analysis, "Correct", "Passing", and "Valid" were mapped to 0, and "Incorrect", "Failing", and "Invalid" were mapped to 1. Based on this definition, we can numerically evaluate the performance of LLMJ for each issue type. 

Following are the data points recorded or calculated on a per-issue basis:

\begin{itemize}
    \item \textbf{Count}: This is the number of files that correspond to each issue ID.
    \item \textbf{Correct/Incorrect Judgments}: This is the number of correct and mistaken evaluations made by the LLMJ on files corresponding to each issue ID, determined by comparing the LLMJ's evaluations against each file's validity according to the above verification system.
    \item \textbf{Accuracy}: Calculated by first determining the number of correct evaluations made by the LLMJ (equal to the count value minus the mistakes value for each issue ID), and dividing that number by the number of files with the same issue ID. The resulting value represents the percentage of correct evaluations made by the LLMJ for each issue ID.
\end{itemize}

Additionally, we conducted a numerical evaluation to assess the overall accuracy and bias.

%The following processes were used to calculate the overall metrics:

\begin{itemize}
    \item \textbf{Overall evaluation accuracy}: Calculated by determining the total number of correct evaluations, and dividing it by the total number of files, regardless of each file's issue ID.
    \item \textbf{Bias}: Calculated by first determining a total bias value. 1 is added to the total for each mistaken evaluation of an invalid file, and 1 is subtracted from the total for each mistaken evaluation of a valid file. The resulting total is then divided by the total number of mistaken evaluations, giving a value in the range [-1, 1].
\end{itemize}
These metrics allowed us to create profiles of multiple different approaches and setups for verifying compiler tests when each approach was subjected to negative probing.
%%%%%%%%%%%%%%%%%%%%%%%%%%%%%%%%%%%%%%%%%%%%%%%%%%%%%%%
%
%                    NEW SECTION
%
%%%%%%%%%%%%%%%%%%%%%%%%%%%%%%%%%%%%%%%%%%%%%%%%%%%%%%%
\section{Analysis of deepseek-coder-33B-instruct as an LLMJ}

This section discusses results from  analyzing deepseek-coder-33B-instruct as an LLMJ. We do so in two parts. 
\begin{itemize}
    \item \textbf{Part One}: We discuss results derived from using the LLMJ by itself through negative probing
    \item \textbf{Part Two}: We discuss results from two different prompting styles for an agent-based LLMJ and a validation pipeline that utilizes an agent-based LLMJ. 
\end{itemize}

\subsection{Results for Part One}

Initial experimentation began with an analysis of the LLMJ technique itself. Two test suites were put together with negative probing to test the LLM against: one suite for OpenMP (containing only C files, due to time constraints), and one suite for OpenACC (containing C, C++, and Fortran files). 
%Due to time constraints on the experiment, we were not able to include compatibility for C++ and Fortran files containing OpenMP.

After assembling the testsuites, we loaded the deepseek-coder-33b-instruct model onto one node on Perlmutter, and use the following prompt for each file as shown in Listing~\ref{lst:directAnalysisPrompt}. 
Because the prompt asks the LLM to directly evaluate the code provided, we called this prompt a direct analysis prompt. 

% Review the following OpenACC/OpenMP code and evaluate it based on the following criteria:

\begin{lstlisting}[numbers=left, numberstyle=\small, stepnumber=1, frame=single, label=lst:prompts, breaklines=true, breakindent=20pt, xleftmargin=20pt, xrightmargin=20pt, caption=Direct Analysis - an Example Prompt,label=lst:directAnalysisPrompt]
Review the following OpenACC/OpenMP code and evaluate it based on the following criteria:

Syntax: Ensure all OpenACC/OpenMP directives and pragmas are syntactically correct.
Directive Appropriateness: Check if the right directives are used for the intended parallel computations.
Clause Correctness: Verify that all clauses within the directives are correctly used according to OpenACC/OpenMP specifications.
Memory Management: Assess the accuracy of data movement between CPU and GPU.
Compliance: Ensure the code adheres to the latest OpenACC specifications and best practices.
Logic: Verify that the logic of the test (e.g. performing the same computation in serial and parallel and comparing) is correct.
Based on these criteria, evaluate the code in a brief summary, then respond with precisely "FINAL JUDGEMENT: correct" (or incorrect).
You MUST include the exact phrase "FINAL JUDGEMENT: correct" in your evaluation if you believe the code is correct. Otherwise, you must include the phrase "FINAL JUDGEMENT: incorrect" in your evalutation.
Here is the code:
{C/C++/Fortran file content}
\end{lstlisting}

% Based on these criteria, evaluate the code in a brief summary, then respond with precisely "FINAL JUDGEMENT: correct" (or incorrect).
% You MUST include the exact phrase "FINAL JUDGEMENT: correct" in your evaluation if you believe the code is correct. Otherwise, you must include the phrase "FINAL JUDGEMENT: incorrect" in your evalutation.\\
% Here is the code:\\
% \{file content\}
% }

The LLM's response and evaluation were then recorded for each file, and we performed an analysis of the data. 
Table~\ref{tab:accissuetype} and Table \ref{tab:mpissuetype} show the per-issue accuracy of deepseek-coder-33B-instruct's evaluations for OpenACC and OpenMP files, respectively. 

\begin{table*}[ht]
 \caption{LLMJ Negative Probing Results for OpenACC} 
\centering
\begin{tabular}{|l|c|c|c|c|}
\hline \multicolumn{1}{|c|}{ OpenACC Issue Type } & \begin{tabular}{c} 
Total \\
Count
\end{tabular} & \begin{tabular}{c} 
Correct \\
Judgments
\end{tabular} & \begin{tabular}{c} 
Incorrect \\
Judgments
\end{tabular} & Accuracy \\
\hline Removed ACC memory allocation /  swapped ACC directive & 203 & 31 & 172 & $15 \%$ \\
\hline Removed an opening bracket & 125 & 15 & 110 & $12 \%$ \\
\hline Added use of undeclared variable & 108 & 16 & 92 & $15 \%$ \\
\hline  Replaced file with randomly-generated non-OpenACC code  & 117 & 94 & 23 & $80 \%$ \\
\hline Removed last bracketed section of code & 114 & 14 & 100 & $12 \%$ \\
\hline No issue & 668 & 586 & 82 & $88 \%$ \\
\hline
\end{tabular}
\label{tab:accissuetype}
\end{table*}

% \begin{figure}[h]
% \centering
% \includegraphics[width=0.50\textwidth]{Images/OpenACC-LLMJ-1-Results-Table.png}
% \caption{LLMJ Negative Probing Results for OpenACC}
% \label{fig:OpenACC-LLMJ-1}
% \end{figure}

\begin{table*}[ht]
 \caption{LLMJ Negative Probing Results for OpenMP} 
\centering
\begin{tabular}{|l|c|c|c|c|}
\hline \multicolumn{1}{|c|}{ OpenMP Issue Type } & \begin{tabular}{c} 
Total \\
Count
\end{tabular} & \begin{tabular}{c} 
Correct \\
Judgments
\end{tabular} & \begin{tabular}{c} 
Incorrect \\
Judgments
\end{tabular} & Accuracy \\
\hline 
Removed OMP memory allocation / swapped OMP directive  & 59 & 28 & 31 & $47 \%$ \\
\hline Removed an opening bracket & 39 & 29 & 10 & $74 \%$ \\
\hline Added use of undeclared variable & 33 & 21 & 12 & $64 \%$ \\
\hline  Replaced file with randomly-generated non-OpenMP code  & 51 & 2 & 49 & $4 \%$ \\
\hline Removed last bracketed section of code & 33 & 11 & 22 & $33 \%$ \\
\hline No issue & 216 & 84 & 132 & $39 \%$ \\
\hline
\end{tabular}
\label{tab:mpissuetype}
\end{table*}

As Table \ref{tab:accissuetype} demonstrates, deepseek-coder-33B-instruct struggled to recognize basic syntax errors and test logic errors in OpenACC files, and was only able to accurately determine whether the test contained any OpenACC directives or routines at all. Meanwhile, Table \ref{tab:mpissuetype} shows that the LLMJ was significantly better at recognizing syntax errors in OpenMP files, while struggling a bit more to recognize OpenMP errors and test logic errors. Notably, the LLMJ was almost entirely incapable of recognizing when a file did not contain any OpenMP at all.

\begin{table}[!ht]
 \caption{LLMJ Overall Negative Probing Results} 
\centering
\begin{tabular}{|l|c|c|}
\hline \multicolumn{1}{|c|}{ Datapoint } & OpenACC & OpenMP \\
\hline Total Count & 1335 & 431 \\
\hline Total Mistakes & 579 & 256 \\
\hline Overall Accuracy & $56.63 \%$ & $40.60 \%$ \\
\hline Bias & 0.717 & -0.031 \\
\hline
\end{tabular}
\label{tab:LLMJNegativeResults}
\end{table}

% \begin{wrapfigure}{h}{0.3\textwidth}
% \centering\includegraphics[width=0.3\textwidth]{Images/LLMJ-1-Results-Table.png}
% \caption{LLMJ Overall Negative Probing Results}
% \label{fig:LLMJ-1}
% \end{wrapfigure}

Table \ref{tab:LLMJNegativeResults} shows the overall performance of deepseek-coder-33B-instruct as a judge for OpenACC as well as OpenMP. Surprisingly, despite OpenMP having existed for a longer period of time, deepseek-coder-33B-instruct demonstrated a higher overall accuracy when evaluating OpenACC files. However, it also exhibited a much higher positive bias for OpenACC than for OpenMP, demonstrating a strong tendency for its mistakes to involve it passing an invalid file.

%%%%%%%%%%%
%
%Past this point, I'm just throwing up words onto the page; the next section needs HEAVY refinement
%
%%%%%%%%%%

\subsection{Results for Part Two}

Based on these results, we concluded that it would be necessary to equip the LLMJ with more tools in order to improve its accuracy. 
We designed the validation pipeline and implemented an agent-based approach for the LLMJ, and created larger testsuites for OpenMP and OpenACC (using C and C++ files from the manually-written testsuites for both).
Many OpenMP offloading compilers do not support all OpenMP features introduced after version 4.5. To reduce the likelihood of this inconsistent feature support affecting our results, we only included files that used OpenMP 4.5 or lower to ensure that the LLVM OpenMP offloading compiler we used would be fully-compliant for all features present.
% In order to prevent a lack of compiler support for newer OpenMP features from affecting our results, we only included files that used OpenMP 4.5 or lower to verify that the compiler we used for OpenMP, LLVM, would be fully-compliant.  CM - shouldnt 4.5 comment be earlier in setup before stage 1 results. 
For OpenACC, we used NVIDIA's HPC SDK nvc compiler. For now, we have experimented mostly with C/C++ files, with an aim to include Fortran files in the near future.  

% We elected to omit Fortran files from our testsuites so as to verify that all modifications made to each file would successfully make that file invalid (we don't know enough about fortran to know if the changes would actually break the file). 
% {\color{red}Need to verify}

%In addition to experimenting with an agent-based approach, we also experimented with another prompting technique, we call this an indirect analysis prompt. 

We theorize that the wording of our direct analysis prompt in Listing \ref{lst:indirect} was causing the LLM to provide results based on examples of code reviews online instead of its knowledge of OpenMP and OpenACC. 
To remedy this, we re-wrote the prompt and instructed the LLM to generate a detailed description of the code provided, and then determine if that description fit the profile of a valid compiler test. In this way, the LLM would be indirectly evaluating the code, so we referred to it as an indirect analysis prompt. With this approach, the LLM would hopefully base its response on its knowledge of OpenMP and OpenACC (when generating a description of the code), and its knowledge of compiler tests (when analyzing the description).
%\newpage
The following shows the indirect anlaysis prompt that we created: 
\begin{lstlisting}[numbers=left, numberstyle=\small, stepnumber=1, frame=single, label=lst:prompts, breaklines=true, breakindent=20pt, xleftmargin=20pt, xrightmargin=20pt, caption=Indirect Analysis - an Example Prompt,label=lst:indirect]
Describe what the below OpenACC/OpenMP program will do when run. Think step by step.
Here is some information about the code to help you; you do not have to compile or run the code yourself.
When the below code is compiled with a OpenACC/OpenMP-compliant compiler, the compiler gives the following outputs:
 Compiler return code: {return code}
 Compiler STDERR: {STDOUT}
 Compiler STDOUT: {STDERR}
When the compiled code is run, it gives the following results:
 Return code: {return code}
 STDOUT: {STDOUT}
 STDERR: {STDERR}
Using this information, describe in full detail how the below code works, what the below code will do when run, and suggest why the below code might have been written this way. 
Then, based on that description, determine whether the described program would be a valid or invalid compiler test for {flavor} compilers. 
You MUST include the exact phrase "FINAL JUDGEMENT: valid" in your final response if you beleive that your description of the below OpenACC/OpenMP code describes a valid compiler test; otherwise, your final response MUST include the exact phrase "FINAL JUDGEMENT: invalid".
Here is the code for you to analyze:  {C/C++/Fortran file}
\end{lstlisting}

For each testsuite, we then compiled, executed, and used both LLMJ prompts to evaluate each file while recording the compilation data, execution data, and evaluations. We ran each file through every stage of the validation pipeline; however, for this experiment, we did not prevent invalid files from continuing through the pipeline. % sounds contradictory to me  CM 
This way, we could gather information about both agent-based LLMJs, and retroactively verify how the entire validation pipeline would have performed on the data by checking the compilation, execution, and evaluation status of each file. \\

To simplify the data analysis: 
\begin{itemize}
    \item LLMJ 1: The agent-based LLMJ that used the direct analysis prompt
    \item LLMJ 2: The agent-based LLMJ that used the indirect analysis prompt 
    \item Pipeline 1: Validation pipeline outputs computed with LLMJ 1's evaluation 
    \item Pipeline 2: Validation pipeline outputs computed with LLMJ 2's evaluation 
\end{itemize}
% we dubbed the agent-based LLMJ that used the direct analysis prompt "LLMJ 1" and dubbed the agent-based LLMJ that used the indirect analysis prompt "LLMJ 2". Similarly, we referred to the validation pipeline outputs computed with LLMJ 1's evaluation "Pipeline 1", and the validation pipeline outputs computed with LLMJ 2's evaluation "Pipeline 2". 
We then compared the performances of the two pipelines against each other, and compared the two agent-based LLMJs against each other and against the non-agent-based LLMJ. 

\begin{table*}[ht]
\caption{Validation Pipeline Results for OpenACC}
\centering
\begin{tabular}{|l|c|c|c|c|c|}
\hline \multicolumn{1}{|c|}{OpenACC Issue Type} & \begin{tabular}{l} 
Total \\ Count
\end{tabular} & \begin{tabular}{c}
Pipeline 1 \\
Correct \\ Evaluations
\end{tabular} & \begin{tabular}{c} 
Pipeline 2 \\ Correct \\ Evaluations
\end{tabular} & \begin{tabular}{c} 
Pipeline 1 \\ Accuracy
\end{tabular} & \begin{tabular}{c} 
Pipeline 2 \\ Accuracy
\end{tabular} \\
\hline 
Removed ACC memory allocation / swapped ACC directive & 272 & 250 & 251 & 92\% & 92\% \\
\hline Removed an opening bracket & 146 & 146 & 146 & $100 \%$ & $100 \%$ \\
\hline Added use of undeclared variable & 151 & 151 & 151 & $100 \%$ & $100 \%$ \\
\hline Replaced file with randomly-generated non-OpenACC code & 146 & 146 & 146 & $100 \%$ & $100 \%$ \\
\hline Removed last bracketed section of code & 176 & 38 & 53 & $22 \%$ & $30 \%$ \\
\hline No issue & 891 & 704 & 627 & 79\% & 70\% \\
\hline
\end{tabular}
\label{tab:OpenACCPipelineResults}
\end{table*}

% \begin{figure}[h]
% \centering
% \includegraphics[width=0.5\textwidth]{Images/OpenACC-Pipeline-Results-Table.png}
% \caption{Pipeline results for OpenACC}
% \label{fig:OpenACC-Pipeline}
% \end{figure}

\begin{table*}[ht]
\centering
\caption{Validation Pipeline Results for OpenMP}
\begin{tabular}{|l|c|c|c|c|c|}
\hline \multicolumn{1}{|c|}{OpenMP Issue Type} & \begin{tabular}{l} Total \\ Count \end{tabular} & \begin{tabular}{c} Pipeline 1 \\ Correct \\ Evaluations \end{tabular} 
& \begin{tabular}{c} Pipeline 2 \\ Correct \\ Evaluations \end{tabular} 
& \begin{tabular}{c} Pipeline 1 \\ Accuracy \end{tabular} 
& \begin{tabular}{c} Pipeline 2 \\ Accuracy \end{tabular} \\
\hline  Removed OMP memory allocation / 
swapped OMP directive & 49 & 47 & 46 & 96\% & 94\% \\
\hline Removed an opening bracket & 28 & 28 & 28 & $100 \%$ & $100 \%$ \\
\hline Added use of undeclared variable & 26 & 26 & 26 & $100 \%$ & $100 \%$ \\
\hline  Replaced file with randomly-generated  non-OpenMP code  & 20 & 14 & 17 & $70 \%$ & $85 \%$ \\
\hline Removed last bracketed section of code & 25 & 23 & 23 & $92 \%$ & $92 \%$ \\
\hline No issue & 148 & 136 & 138 & $92 \%$ & 93\% \\
\hline
\end{tabular}
\label{tab:OpenMPPipelineResults}
\end{table*}

% \begin{figure}[h]
% \centering
% \includegraphics[width=0.5\textwidth]{Images/OpenMP-Pipeline-Results-Table.png}
% \caption{Pipeline results for OpenMP}
% \label{fig:OpenMP-Pipeline}
% \end{figure}

\begin{table}[!ht]
\centering
\caption{Overall Validation Pipeline Results}
\begin{tabular}{|l|c|c|}
\hline \multicolumn{1}{|c|}{Datapoint} & OpenACC & OpenMP \\
\hline Total Count & 1782 & 296 \\
\hline Total Pipeline 1 Mistakes & 347 & 22 \\
\hline Total Pipeline 2 Mistakes & 408 & 18 \\
\hline Overall Pipeline 1 Accuracy & $80.53 \%$ & $92.57 \%$ \\
\hline Overall Pipeline 2 Accuracy & $77.10 \%$ & $93.92 \%$ \\
\hline Pipeline 1 Bias & -0.078 & -0.091 \\
\hline Pipeline 2 Bias & -0.294 & -0.111 \\
\hline
\end{tabular}
\label{tab:PipelineResults}
\end{table}

% \begin{figure}[t]
% \centering
% \includegraphics[width=0.25\textwidth]{Images/Pipeline-Results-Table.png}
% \caption{Overall Pipeline Results}
% \label{fig:Pipeline}
% \end{figure}

Table \ref{tab:OpenACCPipelineResults} shows the results of the two pipelines on the OpenACC testsuite. As can be seen, the two pipelines performed almost identically, though Pipeline 2 demonstrated a higher ability recognize errors in the test's logic, and Pipeline 1 demonstrated a higher ability to recognize when a file contained no errors. Table \ref{tab:OpenMPPipelineResults} also shows a similarity between the two pipelines' performances, though in this case, Pipeline 2 was slightly worse at recognizing OpenMP errors and significantly better at recognizing a lack of OpenMP code.

\indent Table \ref{tab:PipelineResults} shows the overall performance of both pipelines across both OpenACC and OpenMP. Both pipelines were significantly more accurate for OpenMP than for OpenACC, though Pipeline 1 was slightly more accurate than Pipeline 2 for OpenACC and slightly less accurate than Pipeline 2 for OpenMP. For both programming models, both pipelines demonstrated a bias towards restrictiveness, though Pipeline 2 consistently had a stronger bias than Pipeline 1. This demonstrates that for both pipelines, when a mistake does occur, it is more likely to be one of misjudging a valid file rather than one of misjudging an invalid file.

\begin{figure}[h!]
\centering
\includegraphics[width=0.5\textwidth]{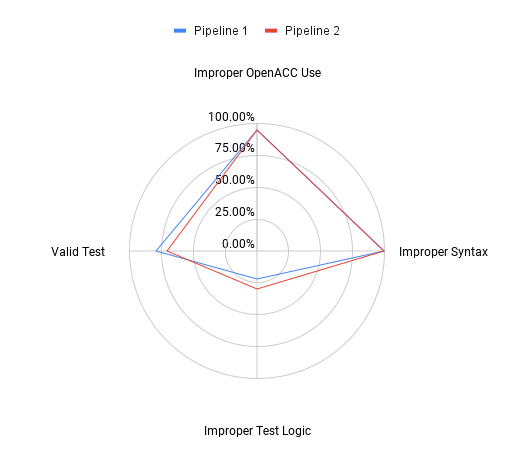}
\caption{A Radar Plot for Validation Pipeline Results for OpenACC}
\label{fig:OpenACC-Radar-Plot}
\end{figure}

\begin{figure}[h!]
\centering
\includegraphics[width=0.5\textwidth]{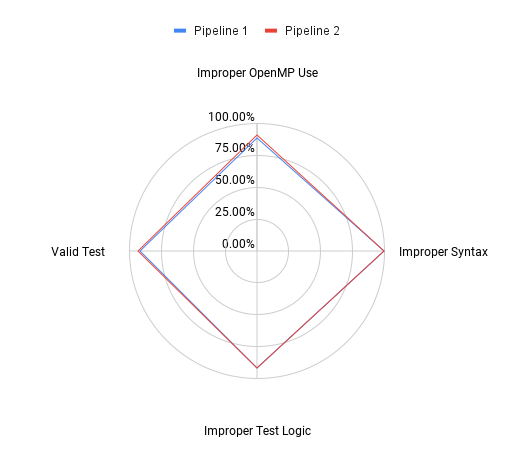}
\caption{A Radar Plot for Validation Pipeline Results for OpenMP}
\label{fig:OpenMP-Radar-Plot}
\end{figure}

\indent Figures \ref{fig:OpenACC-Radar-Plot} and \ref{fig:OpenMP-Radar-Plot} present the accuracy of both pipelines on the four categories of errors introduced into each file, for OpenACC and OpenMP respectively. 
As Figure \ref{fig:OpenMP-Radar-Plot} clearly shows, the performance of both pipelines on OpenMP was nearly identical across all four types of issues, while Pipeline 1 and Pipeline 2 had only slight differences in performance for OpenACC. The radar plots also show the large difference in the pipelines' ability to detect erroneous test logic in OpenMP files versus OpenACC files; however, both pipelines also demonstrated an almost identical ability to detect improper directive use and improper syntax across both OpenACC and OpenMP.

%%%%%%%%%%%%%%%
%               New page
%%%%%%%%%%%%%%%
\begin{table*}[ht]
\centering
\caption{Agent-Based LLMJ Results for OpenACC}
\begin{tabular}{|l|c|c|c|c|c|}
\hline \multicolumn{1}{|c|}{ OpenACC Issue Type } & 
\begin{tabular}{c} Total \\ Count \end{tabular} & 
\begin{tabular}{c} LLMJ 1 \\ Correct \\ Evaluations \end{tabular} & 
\begin{tabular}{c} LLMJ 2 \\ Correct \\ Evaluations \end{tabular} & 
\begin{tabular}{c} LLMJ 1 \\ Accuracy \end{tabular} & 
\begin{tabular}{c} LLMJ 2 \\ Accuracy \end{tabular} \\
\hline  Removed ACC memory allocation / swapped ACC directive & 272 & 182 & 224 & $67 \%$ & $82 \%$ \\
\hline  Removed an opening bracket & 146 & 111 & 81 & $76 \%$ & $55 \%$ \\
\hline Added use of undeclared variable  & 151 & 128 & 126 & $85 \%$ & $83 \%$ \\
\hline  Replaced file with randomly-generated  non-OpenACC code & 146 & 142 & 146 & $97 \%$ & $100 \%$ \\
\hline Removed last bracketed section of code & 176 & 26 & 47 & $15 \%$ & $27 \%$ \\
\hline No issue & 891 & 819 & 701 & $92 \%$ & $79 \%$ \\
\hline
\end{tabular}
\label{tab:OpenACC-LLMJ-2}
\end{table*}

% \begin{figure}[h]
% \centering
% \includegraphics[width=0.5\textwidth]{Images/OpenACC-LLMJ-2-Results-Table.png}
% \caption{Pipeline results for OpenACC}
% \label{fig:OpenACC-LLMJ-2}
% \end{figure}

\begin{table*}[ht]
\centering
\caption{Agent-Based LLMJ Results for OpenMP}
\begin{tabular}{|l|c|c|c|c|c|}
\hline \multicolumn{1}{|c|}{OpenMP Issue Type} & 
\begin{tabular}{l} Total \\ Count \end{tabular} & 
\begin{tabular}{c} LLMJ 1 \\ Correct \\ Evaluations \end{tabular} & 
\begin{tabular}{c} LLMJ 2 \\ Correct \\ Evaluations \end{tabular} & 
\begin{tabular}{c} LLMJ 1 \\ Accuracy \end{tabular} & 
\begin{tabular}{c} LLMJ 2 \\ Accuracy \end{tabular} \\
\hline  Removed OMP memory allocation /  swapped OMP directive  & 49 & 23 & 22 & $47 \%$ & $45 \%$ \\
\hline Removed an opening bracket & 28 & 16 & 13 & $57 \%$ & $46 \%$ \\
\hline Added use of undeclared variable & 26 & 18 & 15 & $69 \%$ & $58 \%$ \\
\hline  Replaced file with randomly-generated non-OpenMP code  & 20 & 13 & 17 & $65 \%$ & $85 \%$ \\
\hline Removed last bracketed section of code & 25 & 18 & 12 & $72 \%$ & $48 \%$ \\
\hline No issue & 148 & 137 & 142 & 93\% & 96\% \\
\hline
\end{tabular}
\label{tab:OpenMP-LLMJ-2}
\end{table*}

% \begin{figure}[h]
% \centering
% \includegraphics[width=0.5\textwidth]{Images/OpenMP-LLMJ-2-Results-Table.png}
% \caption{Pipeline results for OpenMP}
% \label{fig:OpenMP-LLMJ-2}
% \end{figure}

\begin{table}[!ht]
\caption{Overall Agent-Based LLMJ Results}
\centering
\begin{tabular}{|l|c|c|}
\hline \multicolumn{1}{|c|}{Datapoint} & OpenACC & OpenMP \\
\hline Total Count & 1782 & 296 \\
\hline Total LLMJ 1 Mistakes & 374 & 71 \\
\hline Total LLMJ 2 Mistakes & 457 & 75 \\
\hline Overall LLMJ 1 Accuracy & 79.01\% & 76.01\% \\
\hline Overall LLMJ 2 Accuracy & 74.35\% & 74.66\% \\
\hline LLMJ 1 Bias & 0.615 & 0.690 \\
\hline LLMJ 2 Bias & 0.168 & 0.840 \\
\hline
\end{tabular}
\label{tab:LLMJ-2}
\end{table}

% \begin{figure}[t!]
% \centering
% \includegraphics[width=0.25\textwidth]{Images/LLMJ-2-Results-Table.png}
% \caption{Overall Agent-Based LLMJ Results}
% \label{fig:LLMJ-2}
% \end{figure}

\begin{figure}[h!]
\centering
\includegraphics[width=0.5\textwidth]{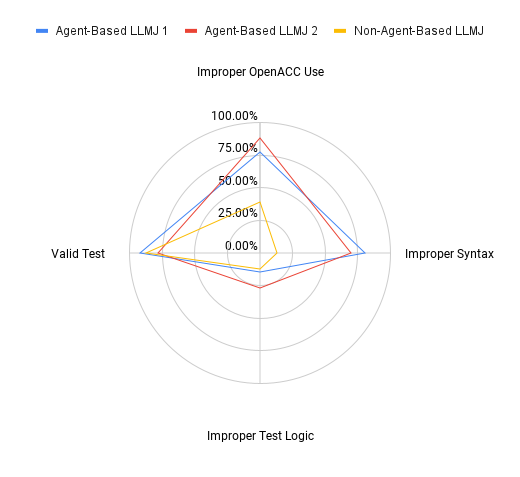}
\caption{A Radar Plot for LLMJ Results for OpenACC}
\label{fig:OpenACC-LLMJ-Plot}
\end{figure}

\begin{figure}[h!]
\centering
\includegraphics[width=0.5\textwidth]{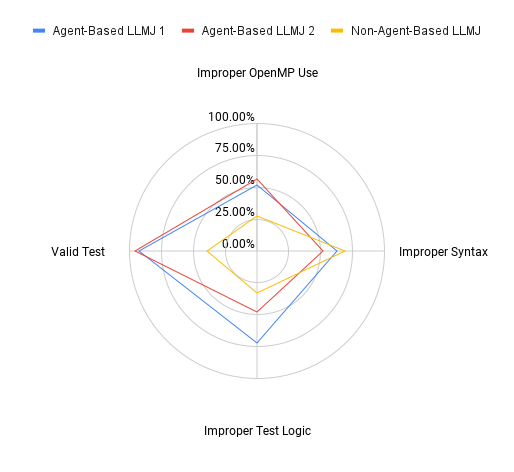}
\caption{A Radar Plot for LLMJ Results for OpenMP}
\label{fig:OpenMP-LLMJ-Plot}
\end{figure}

\indent Table \ref{tab:OpenACC-LLMJ-2} shows the results of the two agent-based LLMJs on the OpenACC testsuite. In this case, the two LLMJs' performances varied much more, with LLMJ 1 demonstrating a superior ability to identify missing syntax errors and to recognize valid code, and LLMJ 2 demonstrating a superior ability to detect OpenACC errors, a lack of OpenACC code, and errors in test logic. Table \ref{tab:OpenMP-LLMJ-2}, which shows the performance of both LLMJs on OpenMP, also demonstrates more variance between the two LLMJs. LLMJ 1 exhibited a higher accuracy for recognizing OpenMP errors, syntax errors, and test logic error, while LLMJ 2 was better-equipped for recognizing a lack of OpenMP code and valid codes. 

%%%   Below segment is only here for formatting purposes
\indent Table \ref{tab:LLMJ-2} shows the overall performance of both LLMJs. For both OpenACC and OpenMP, LLMJ 1 demonstrated a higher overall accuracy than LLMJ 2 and performed slightly better on OpenACC than OpenMP, while LLMJ 2 demonstrated a roughly equal level of accuracy between 
\\\\
OpenACC and OpenMP. LLMJ 1 also exhibited a consistently strong positive bias, while LLMJ 2 had a much smaller positive bias for OpenACC and a much larger positive bias for OpenMP. In all cases, the agent-based LLMs exhibited a tendency towards passing invalid files as opposed to failing valid files. 
Compared to the non-agent-based LLMJ, both agent-based LLMJs exhibited drastically higher overall accuracy, and both exhibited a smaller positive bias for OpenACC. 

\indent Figures \ref{fig:OpenACC-LLMJ-Plot} and \ref{fig:OpenMP-LLMJ-Plot} present the accuracy of all three LLMJs for OpenACC and OpenMP, respectively. In almost all categories, the agent-based LLMJs outperformed the non-agent-based LLMJ, with the exception of valid test recognition for OpenACC (where the non-agent-based LLMJ outperformed LLMJ 2), and improper syntax recognition for OpenMP (where the non-agent-based LLMJ outperformed both agent-based LLMJs). LLMJ 2 consistently demonstrated a higher accuracy in recognizing improper directive usage than LLMJ 1, while LLMJ 1 exhibited a better recognition of improper syntax than LLMJ 2. Both agent-based LLMJs were also consistently able to recognize valid tests with a high degree of accuracy, with LLMJ 1 slightly outperforming LLMJ 2 for OpenACC.

\section{Conclusion}
In this paper, we explore ways to assess the capability of LLM-as-a-Judge. We employ different techniques such as negative probing and agent-based approach along with prompts to understand how the LLM evaluates the codes. 
Our results indicate that utilizing an agent-based prompting approach and setting up a validation pipeline structure significantly increased the quality of DeepSeek Coder's evaluations of tests used to validate compiler implementations of directive-based programming models. As part of our future work, we will incorporate fortran code into our testing to ensure more comprehensive data collection and probing. We will also be exploring the automation of compiler test generation based on lessons learnt from this work.

% conference papers do not normally have an appendix

% use section* for acknowledgment
\section*{Acknowledgment}
The authors are very grateful to OpenACC for supporting this work. This research used resources NERSC, a U.S. DOE Office of Science User Facility located at LBNL, operated under Contract No. DE-AC02-05CH11231 using NERSC ERCAP0029463. This material is also based upon work supported by the U.S. DOE under Contract DE-FOA-0003177, S4PST: Next Generation Science Software Technologies Project.

% trigger a \newpage just before the given reference
% number - used to balance the columns on the last page
% adjust value as needed - may need to be readjusted if
% the document is modified later
%\IEEEtriggeratref{8}
% The "triggered" command can be changed if desired:
%\IEEEtriggercmd{\enlargethispage{-5in}}

% references section

% can use a bibliography generated by BibTeX as a .bbl file
% BibTeX documentation can be easily obtained at:
% http://mirror.ctan.org/biblio/bibtex/contrib/doc/
% The IEEEtran BibTeX style support page is at:
% http://www.michaelshell.org/tex/ieeetran/bibtex/
\bibliographystyle{IEEEtran}
% argument is your BibTeX string definitions and bibliography database(s)
% \bibliography{IEEEabrv,main}
\bibliography{main}

% that's all folks
\end{document}